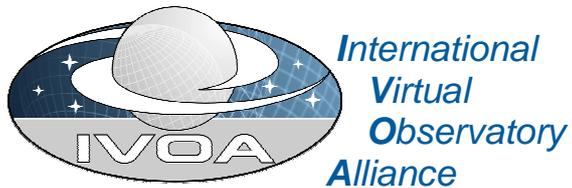

# IVOA Document Standards Version 1.2

## IVOA Recommendation  13 April 2010

## Abstract
This document describes the types of official IVOA documents and the process by which documents are advanced from Working Drafts to formal Recommendations.

## Status of this document

This document has been produced by the IVOA Standing Committee on Standards and Process.

It has been reviewed by IVOA Members and other interested parties, and has been endorsed by the IVOA Executive Committee as an IVOA Recommendation. It is a stable document and may be used as reference material or cited as a normative reference from another document. IVOA's role in making the Recommendation is to draw attention to the specification and to promote its widespread deployment. This enhances the functionality and interoperability inside the Astronomical Community.

A list of current IVOA Recommendations and other technical documents can be found at http://www.ivoa.net/Documents/.

## Acknowledgments
This document is based on the W3C documentation standards, but has been adapted for the IVOA.

# Contents

# 1  Document types

The IVOA publishes two types of documents:

**Recommendation track documents.**  These are specifications, guidelines, etc. produced by Working Groups. Documents on the Recommendation track may progress from Working Draft (WD) to Proposed Recommendation (PR) and finally to Recommendation (REC).

**IVOA Notes.**  An IVOA Note is a dated, public record of an idea, comment, or document. Authorship of a Note may vary greatly (e.g., by an IVOA Working Group, by an IVOA member, etc.).

All public documents are available at the IVOA document repository Web site, http://www.ivoa.net/Documents/.  IVOA will make every effort to make archival documents indefinitely available at their original address in their original form.

The IVOA Executive Committee appoints a Documentation Coordinator (DC) who oversees the document collection and assures that documents conform to these guidelines.

The DC may reformat, rename, or renumber documents so as to conform to changes in IVOA practice (e.g., changes to document styles or the "Status of this Document" section).

Each public document must clearly indicate whether it is a Note, Working Draft (WD), Proposed Recommendation (PR), or Recommendation (REC).

The primary language for IVOA documents is English.

## 1.1 Status

Each document must include a section about the status of the document. The status section should explain why IVOA has published the document, whether or not it is part of the Recommendation track, who developed it, where to send comments about it, whether implementation experience is being sought, any significant changes from the previous version, and any other relevant metadata.

The status section of a Working Draft must set expectations about the stability of the work (e.g., that it may be superseded, obsoleted, or dropped at any time, that it should not be cited as other than a work in progress, etc.) and must indicate how much consensus within IVOA there is about the Working Draft (e.g., no consensus, consensus among the Working Group participants, etc.).

The status section of a Note must indicate the level of endorsement within or by IVOA for the material in the Note, and set expectations about future commitments from IVOA to pursue the topics covered by the Note or to respond to comments about the Note.

## 1.2 Naming and version numbering conventions

IVOA document names have five components:

1. A document type code: NOTE, WD (Working Draft), PR (Proposed Recommendation), or REC (Recommendation).
2. A concise name, which should be a reasonable condensation of the document title.
3. A version number of the form I.J, where I and J are integers 0, 1, 2, ... 9, 10, 11, ... .
4. A date. The date is the GMT date on which the current version of the document was produced, in the format YYYYMMDD. (This does not allow for multiple versions of a document to be released within one 24-hour period, but this should not be a major problem.)
5. An extension (.html, .pdf, .doc, etc.) that follows MIME type conventions.

The first four components are concatenated, separated by hyphens.

Version numbers follow these guidelines:
- The number to the left of the (first) decimal point starts with 0 for documents that are being discussed within a Working Group prior to publication for IVOA-wide review. The number increments to 1 for the first public version, and to 2, 3, ..., for subsequent versions that are not backward compatible and/or require substantial revisions to implementations.
- The number to the right of the decimal point is an integer counter, beginning with 0 and increasing in simple cardinal order (0, 1, 2, ... 9, 10, 11, ...). This number does not track every revision to a document, but rather, denotes a logical version or conceptually consistent view. This number should only be incremented when there are significant and substantial changes to text but few (minor) or no changes required of implementations. The version number normally remains fixed as a document is promoted from Working Draft to Proposed Recommendation to Recommendation, with editorial revisions indicated by the change of date.
- After a document reaches Recommendation status, subsequent revisions retrace the promotion process. Changes that are backward compatible result in increments in the number to the right of the decimal place (1.1, 1.2, ...). Changes that are not backward compatible require an increment of the number of the left of the decimal place (2.0), with subsequent backward compatible revisions following the same pattern (2.1, 2.2, ...).

The final published and approved Recommendation retains the date on the title page of the document, but the date is removed from the document filename in order to simplify reference to the document.

The following examples show a typical name and numbering progression for a sample document.

| | |
|---|---|
| NOTE-MyNewIdea-1.0-20081001.pdf | (initial idea) |
| WD-ConciseName-0.1-20081225.pdf | (first Working Draft, in WG) |
| WD-ConciseName-0.1-20081231.pdf | (revised 6 days later) |
| WD-ConciseName-0.2-20090115.pdf | (text revised substantially) |
| WD-ConciseName-0.2-20090201.pdf | (final version in WG before PR) |
| WD-ConciseName-1.0-20090301.pdf | (published first version) |
| PR-ConciseName-1.0-20090501.pdf | (promoted to PR) |
| PR-ConciseName-1.0-20090615.pdf | (updated after RFC) |
| PR-ConciseName-1.0-20090801.pdf | (updated after TCG review) |
| REC-ConciseName-1.0.pdf | (accepted as REC; date, e.g., 20090901 appears on title page) |
| | |
| WD-ConciseName-1.1-20100628.pdf | (first update to WD in WG; does not affect software) |
| WD-ConciseName-1.1-2010715.pdf | (revised text) |
| WD-ConciseName-1.1-2010801.pdf | (revised text) |
| PR-ConciseName-1.1-20100815.pdf | (promoted to PR) |
| PR-ConciseName-1.1-201000915.pdf | (updated after RFC) |
| PR-ConciseName-1.1-20101001.pdf | (updated after TCG review) |
| REC-ConciseName-1.1.pdf | (accepted as REC) |
| | |
| WD-ConciseName-2.0-20110628.pdf | (major update to WD in WG; does affect software) |
| WD-ConciseName-2.0-2011715.pdf | (revised text) |
| WD-ConciseName-2.0-2011801.pdf | (revised text) |
| PR-ConciseName-2.0-20110815.pdf | (promoted to PR) |
| PR-ConciseName-2.0-201100915.pdf | (updated after RFC) |
| PR-ConciseName-2.0-20111001.pdf | (updated after TCG review) |
| REC-ConciseName-2.0.pdf | (accepted as REC) |

Names will be reviewed and may be modified by the Document Coordinator to be consistent with these conventions. All versions 1.0 and higher are stored in the IVOA Document Repository.[1]

## 1.3  Format

The standard format for IVOA documents is PDF, though any document preparation tools may be used that allow for the publication of PDF and that retain the standard formatting elements and style. Document templates are provided for MSWord and HTML at http://www.ivoa.net/Documents/templates/. The document source in its original format should also be submitted and retained in the IVOA document collection.

---

[1] http://www.ivoa.net/Documents/

## 1.4 How to publish a document

Documents are entered into the IVOA document collection by the Document Coordinator in response to a request from a Working Group chair or the person primarily responsible for editing a particular document. A request is initiated by filling out an [on-line form](http://www.ivoa.net/bin/up.cgi)[2] and uploading the document. It may be necessary to *tar* or *zip* the document because only a single file can be uploaded at a time. Absolute path names must be avoided when packaging it up as well as when creating internal links within the document.

Since the upload mechanism presents a security risk it cannot be guaranteed to be available at all times and will only accept files up to a certain size. In general such limitations are beyond the control of the DC and altered firewall settings may interfere unexpectedly. In such cases it is necessary to agree with the DC on different means of electronic transfer.

## 1.5 Supplementary resources

The Document Coordinator maintains a repository of supplementary resources, such as XML schema, RDF vocabulary definitions, and WSDL files. Developers and any type of validation system/service should use these in preference to copies stored elsewhere. There is, however, no requirement to use them if a different implementation yields compliance with a given standard. Such additional items are considered part of the implementation but not part of the standard itself. Standards document authors should, however, reference them as informative appendices if applicable and seek consistency. At the same time, authors of auxiliary files should include comments stating which standards and versions thereof they support.

## 2 Standards process

The IVOA standards process is used to build consensus around a Virtual Observatory technology, both within IVOA and in the VO community as a whole. IVOA Working Drafts become Recommendations by following this process. The labels that describe increasing levels of maturity and consensus in the standards process are:

**Note.** An IVOA Note is a dated, public record of an idea, comment, practice, experience, insight, advice, guideline, or policy. Authorship of a Note may vary greatly (e.g., by an IVOA Working Group, by an IVOA member, etc.). In some circumstances a Note may be the basis for a Working Draft, but typically Notes are used to describe items relevant to the IVOA other than descriptions of standards or protocols.

**Working Draft.** A document begins as a Working Draft. A Working Draft is a chartered work item of a Working Group and generally represents work in progress and a commitment by IVOA to pursue work in a particular area. The label "Working Draft" does not imply that there is consensus within IVOA about the document.

---

[2] http://www.ivoa.net/bin/up.cgi

**Proposed Recommendation.**  A Proposed Recommendation is believed to meet the relevant requirements of the Working Group's charter and any accompanying requirements documents, to represent sufficient implementation experience, and to adequately address dependencies from the IVOA technical community and comments from previous reviewers.

**IVOA Recommendation.**  An IVOA Recommendation is a document that is the end result of extensive consensus-building within the IVOA about a particular technology or policy.  IVOA considers that the ideas or technology specified by a Recommendation are appropriate for widespread deployment and promote [IVOA's mission]().[3]

Generally, Working Groups create Working Drafts with the intent of advancing them through the standards process. However, publication of a document at one maturity level does not guarantee that it will advance to the next. Some documents may be dropped as active work or may be subsumed by other documents.  If, at any maturity level of the standards process, work on a document ceases (e.g., because a Working Group or activity closes, or because the work is subsumed by another document), a final version of the document should be issued with the status section noting that work on this document has concluded, and for what rationale, and with links provided to relevant follow-on documents.  Any time a document advances to a higher maturity level, the announcement of the transition must indicate any formal objections.  If, at any maturity level prior to Recommendation, review comments or implementation experience result in substantive changes to a document, the document should be returned to Working Draft for further work.  The relationship between Working Drafts, Proposed Recommendations, and Recommendations is shown in the figure below.

## 2.1  Working Draft (WD)

IVOA official documents begin as Working Drafts.  Working Drafts are the purview of a Working Group.  Working Drafts may undergo numerous revisions during their development.  During this volatile phase Working Drafts are not included in the formal IVOA document collection, but rather are maintained by the responsible working group in its area of the IVOA TWiki.

**Entrance criteria.**  A Working Draft is published at the discretion of a Working Group once the WG is satisfied that the document is sufficiently developed to merit broader exposure and feedback.  Publication of a Working Draft is not an assertion of consensus, of endorsement, or of technical and editorial quality. Consensus is not a prerequisite for approval to publish; the Working Group may request publication of a Working Draft even if it is unstable and does not meet all Working Group requirements.  Working Drafts are subject to review by the document coordinator for compliance to these guidelines.

**Ongoing work.**  Once a Working Draft has been published, the Working Group should continue to develop it by encouraging review and feedback within and outside of IVOA.

**Next maturity level.**  After a suitable review and trial period, the chair of the Working Group may promote the Working Draft to a Proposed Recommendation.  Such advancement should occur only when the chair of the Working Group is satisfied that consensus has been reached, and more formal and extensive review is now warranted. Advancement to Proposed Recommendation implies:
1. The Working Group has fulfilled the relevant requirements of the Working Group charter and those of any accompanying requirements documents.
2. The Working Group has formally addressed issues raised during the development and review process (possibly modifying the document).
3. The Working Group has reported all formal objections.
4. Each feature of the Working Draft has been implemented. The Working Group should be able to demonstrate two interoperable implementations of each feature, and validation tools should be available. If the chair of the Working Group believes that broader review is critical, the chair may advance the document to Proposed Recommendation even without adequate implementation experience. In this case, the document status section should indicate why the chair promoted the document directly to Proposed Recommendation. A report describing the implementations or any associated validation tools should be published as a Note, or should be documented as part of the Request for Comments (see below).

---

[3] http://www.ivoa.net/pub/info/index.html

## 2.2  Proposed Recommendation (PR)

**Entrance criteria.**  Proposed Recommendations are published by the chair of a Working Group following the criteria described above.  Proposed Recommendations are considered to be technically mature and ready for wide review.

**Ongoing work.**   The Working Group should continue to encourage review and feedback within and outside of IVOA.

**Next maturity level.**  After a publication period of at least two weeks, the chair of the Working Group that developed the Proposed Recommendation may call for a formal Request for Comments (RFC).  The RFC is sent to the widest possible IVOA distribution lists ([interop@ivoa.net](interop@ivoa.net)) and published by adding a link to the RFC on the IVOA document repository web page.  Distribution of the RFC initiates a four-week public review period.  All comments submitted during this review period must be posted publicly and responded to publicly.  If the review identifies significant deficiencies in the document, such that revisions must be undertaken beyond minor editorial changes or where revisions require changes to software based on the document, the document must be returned to the Working Draft status.  Members of the Technical Coordination Group (TCG), composed of the chairs and vice chairs of other Working Groups and Interest Groups, must examine Proposed Recommendations during the RFC period and post comments in the public record.  Comments from TCG members may be no more than "read and approved," or "no dependency" but if TCG members have significant concerns it is during the RFC period that these must be documented.  It is sufficient to have one input per WG and IG.

Following the RFC period, the WG Chair may issue a revised version of the document that takes into account the comments received during the RFC.  (Such revisions must be minor in nature, or else the document must return to Working Draft status.)  The TCG then has four weeks to make a final review of the document and the public record of comments and responses as a final check for interface problems or compatibility concerns with the standards developed by other Working Groups.  TCG members are required to note their approval of and/or comments about the document on the RFC public comment website.  It is sufficient to have one input per WG and IG.

PRs being brought forward for promotion to REC should, when applicable, have at least two interoperable implementations.  In its final review the TCG may agree to waive this requirement if there are extenuating circumstances.  The chair of the TCG, working in consultation with the chair of the Working Group responsible for the PR, then makes a final summary recommendation, and the chair of the TCG forwards this recommendation to the Executive Committee for review and approval. If the TCG does not agree to waive the requirement regarding interoperable implementations, but there are otherwise no outstanding issues or unresolved problems, the final decision on promotion of the PR to REC rests with the Executive Committee.  If the Executive Committee is satisfied that all comments and concerns have been properly taken into account, they promote the document to a Recommendation.

## 2.3  Recommendation (REC)

**Entrance criteria.**  Recommendations are published by the IVOA Executive Committee following the criteria described above.  Recommendations are the final form of IVOA documents and constitute an IVOA Standard.

**Ongoing work.**   Recommendations may need to be revised and extended as time goes on.  Significant revisions of Recommendations must proceed through the Working Draft and Proposed Recommendation phases.  A significant revision is any revision that requires changes in software based on the document.

**Next maturity level.**  A Recommendation is the highest level of maturity for an IVOA document.  The IVOA Executive Committee may propose that Recommendations be endorsed as standards by the International Astronomical Union, working through IAU Commission 5.

## 2.4  Document promotion process summary

The IVOA document promotion process is summarized in graphical form in the figure below.

1. Working Group prepares Working Draft (version ≥1.0) and submits to Document Coordinator for posting in the IVOA document collection.
2. Working Group reviews the Working Draft. Two reference implementations of any associated software are expected, as well as provision of validation tools.
3. The Chair of the Working Group, with consent of the WG, promotes the document to a Proposed Recommendation and submits it to the Document Coordinator for posting in the IVOA document collection.
4. After a minimum publication period of two weeks, the Chair of the Working Group issues a formal Request for Comments (RFC) to the e-mail distribution list interop@ivoa.net. The RFC and all comments must be logged on a TWiki page whose URL is given in the RFC. A minimum comment period of 4 weeks must be allowed. The chairs or vice chairs of other Working Groups and Interest Groups are required to examine Proposed Recommendations during the RFC period and to post comments in the public record.
5. The Working Group Chair responds to comments on the TWiki page. If comments lead to significant changes to the document, the status reverts to Working Draft (back to Step 1).

6. If comments are addressed to the satisfaction of the WG Chair and WG members, the WG Chair requests a final review, to be completed within 4 weeks, by the Technical Coordination Group, and they add their final comments to the RFC record. The chair of the Technical Coordination Group, working in consultation with the chair of the Working Group responsible for the PR, then makes a final summary recommendation and the chair of the Technical Coordination Group submits the PR to the Executive Committee for approval.
7. The Executive Committee is polled by the IVOA Chair to ascertain if there is consensus for promotion to Recommendation.
8. If yes, the IVOA Chair reports on approval to the TCG and WG Chairs and asks the Document Coordinator to update the document status to Recommendation. If no, the concerns of the IVOA Executive need to be resolved and a new poll taken, or if serious revisions are required, the document would revert to Step 1.
9. The IVOA Executive may propose to the IAU Commission 5 that IVOA Recommendations be endorsed as IAU Standards.

IVOA Document Standards Process

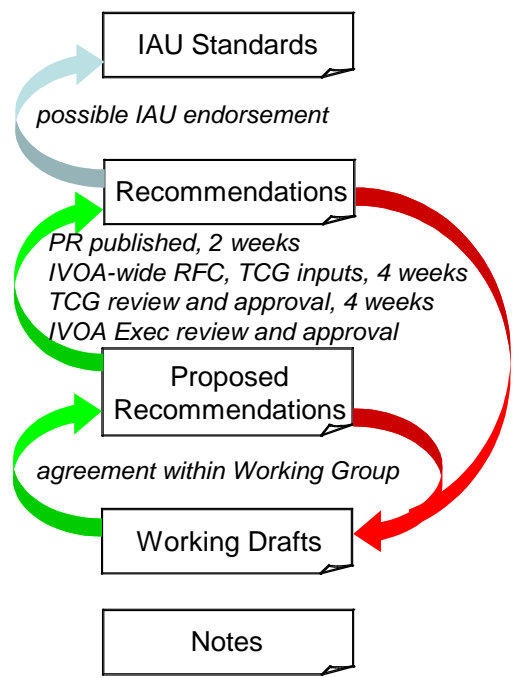

## 3 The document collection

The IVOA document collection is the primary source for IVOA documents. IVOA users, especially from outside the core collaboration, should always be directed to the document collection rather than be sent private copies of documents.

The IVOA document collection is organized so as to lead readers most naturally to the current versions of all documents in all document classes. A document archive is also maintained so that previous published versions of documents remain available (but not pre-published versions of Working Drafts, though working groups may opt to retain this on the TWiki). As an aid to readers, the document collection web site has a link to the Community TWiki area,[4] where Working Drafts in-progress may be found, but links will not be provided to individual pre-publication Working Drafts.

# 4 Changes from previous versions

From 1.1 to 1.2
- In Section 1, added explicit reference to the URL for the IVOA document repository.
- In Section 1, added "The DC may reformat, *rename, or renumber documents*" to allow for standardization in naming and numbering of IVOA documents.
- Changed "technical report" to "document" throughout.
- Rewrote Section 1.2 to include new naming and numbering scheme using .pdf file extensions to be consistent with change to Section 1.3.
- In Section 1.3, changed the required format for documents to PDF and added requirement to submit and store document in its original format.
- Added Section 1.4 "How to Publish a Document" from the IVOA Note, *Guidelines and Procedures for IVOA Document Standards Management* V1.0.
- Added Section 1.5 "Supplementary Resources" from the IVOA Note, *Guidelines and Procedures for IVOA Document Standards Management* V1.0.
- In Section 2, added paragraph to clarify the function of Notes.
- In Section 2.1, added introductory paragraph describing early phases of WD development from the IVOA Note, *Guidelines and Procedures for IVOA Document Standards Management* V1.0, and clarified entrance criteria for Working Draft.
- In Section 2.1, added text describing documentation of interoperable implementations.
- In Section 2.2, clarified how RFCs are published through the IVOA document repository web page.
- In Sections 2.2 and 2.4, updated the role of the TCG in the review of documents during the RFC period. Section 2.2 includes text concerning the requirement for at least two interoperable implementations and under what circumstances this requirement can be waived.
- In Section 2.4, updated the diagram to show more clearly the TCG participation.
- Added Section 3 "The document collection" with text from Sections 6 and 7 of the IVOA Note, *Guidelines and Procedures for IVOA Document Standards Management* V1.0.
- Updated table of contents.

From 1.0 to 1.1
- The role of the Technical Coordination Group (which comprises the Working Group chairs and deputies and Interest Group chairs and deputies) has been made explicit in Section 2.2 describing the RFC process.
- A summary of the process has been added in Section 2.4 and the figure showing the process has been moved into this new section.
- The figure has been updated to reflect the TCG review and to show that Recommendations *may* be referred to the IAU as appropriate.
- Added Appendix with sample text for describing the Status of a document.

---

[4] http://www.ivoa.net/twiki/bin/view/IVOA/WebHome

# Appendix: Recommended Text for Document Status

The following text examples may be used as templates in the Status portion of the document.

*Note*
This is an IVOA Note expressing suggestions from and opinions of the authors. It is intended to share best practices, possible approaches, or other perspectives on interoperability with the Virtual Observatory. It should not be referenced or otherwise interpreted as a standard specification.

*Working Draft*
This is an IVOA Working Draft for review by IVOA members and other interested parties. It is a draft document and may be updated, replaced, or obsoleted by other documents at any time. It is inappropriate to use IVOA Working Drafts as reference materials or to cite them as other than "work in progress".

*Proposed Recommendation*
This is an IVOA Proposed Recommendation made available for public review. It is appropriate to reference this document only as a recommended standard that is under review and which may be changed before it is accepted as a full Recommendation.

*Recommendation*
This document has been produced by the IVOA [working group name] Working Group. It has been reviewed by IVOA Members and other interested parties, and has been endorsed by the IVOA Executive Committee as an IVOA Recommendation. It is a stable document and may be used as reference material or cited as a normative reference from another document. IVOA's role in making the Recommendation is to draw attention to the specification and to promote its widespread deployment. This enhances the functionality and interoperability inside the Astronomical Community.